\documentclass[conference]{IEEEtran}
\IEEEoverridecommandlockouts
\usepackage[T1]{fontenc}
\usepackage{amsmath,amssymb,amsthm,bm}
\usepackage{graphicx,booktabs,array}
\usepackage{enumitem}
\usepackage{cite}
\usepackage{flafter}
\usepackage{needspace}
\usepackage[hidelinks]{hyperref}
\hypersetup{pdftitle={6G Uplink Waveform and Rank Selection under Distributed Reception},pdfkeywords={6G uplink, DFT-s-OFDM, CP-OFDM, distributed reception, waveform switching, rank selection}}
\newtheorem{theorem}{Theorem}
\newtheorem{proposition}{Proposition}
\newtheorem{corollary}{Corollary}
\newenvironment{NewProof}{{\noindent\itshape Proof.}}{\hfill$\blacksquare$\par}
\newcommand{\E}{\mathbb E}

\newcommand{\argmax}{\operatorname*{arg\,max}}
\title{6G Uplink Waveform and Rank Selection under Distributed Reception}
\author{Yewen Cao, Fen Hou, Yulin Shao%
\thanks{Y. Cao and F. Hou are with the State Key Laboratory of Internet of Things for Smart City, University of Macau, Macao, China.}
\thanks{Y. Shao is with the Department of Electrical and Computer Engineering, The University of Hong Kong, Hong Kong, China.}
}

\begin{document}
\bstctlcite{WSDR:reference-style}
\maketitle
\begin{abstract}
The development of 6G uplink is expanding waveform and spatial transmission options, while distributed reception makes their relative merits depend on the spatial and frequency structure of the aggregate channel. We study the joint selection of cyclic prefix orthogonal frequency-division multiplexing and discrete Fourier transform spread OFDM at rank 1 and noncoherent rank 2 under common user equipment power constraints and centralized linear minimum mean-square error reception. We derive waveform switching thresholds at both ranks and a rank selection margin, establishing conditions for selecting among the four modes based on available transmit power, DFT despreading loss, and spatial multiplexing gain. We then characterize how port gains and correlations at individual access points (APs) form aggregate spatial separation and influence mode selection. At fixed port gains and available transmit powers, improving spatial separation without reducing it on any subcarrier increases the rank selection margin, while its effect on the preferred waveform within rank 2 depends on the frequency distribution of the improvement. Numerical results show that different AP deployments can favor different waveforms and ranks, even with the same receive antenna budget and comparable aggregate channel gains, including waveform switches at a fixed preferred rank.
\end{abstract}
\section{Introduction}\label{sec:intro}

Recent 3GPP developments have expanded the options for uplink transmission. New Radio (NR) Release 18 introduced dynamic switching between cyclic-prefix orthogonal frequency-division multiplexing (CP-OFDM) and discrete Fourier transform spread OFDM (DFT-s-OFDM) \cite{tr21918}, while the Release 20 study on the 6G radio interface considers rank-2 DFT-s-OFDM \cite{ran125wave}. 
These developments motivate the joint selection of waveform and transmission rank. Waveform design influences peak-to-average power ratio and hence the available transmit power \cite{shaoliew2020,liewshao2020}, while transmission rank affects spatial multiplexing and receiver equalization \cite{zhao2015}.

This dependence becomes more intricate under distributed reception. Geographically separated access points (APs) observe different channel gains and spatial correlations, and their signals are jointly processed at a central receiver \cite{bjornson2020,wang2022,cao2026networked}. The resulting aggregate channel determines not only how well the transmitted streams can be separated, but also how their spatial separation varies across subcarriers. These properties affect waveform and rank selection differently. Spatial separation governs the benefit of rank-2 transmission, whereas its variation across frequency also affects the despreading loss of DFT-s-OFDM. Consequently, different AP deployments may favor different waveforms and transmission ranks, even with the same receive-antenna budget and comparable aggregate channel gains.

\begin{figure}[!t]
\centering\includegraphics[width=0.95\columnwidth]{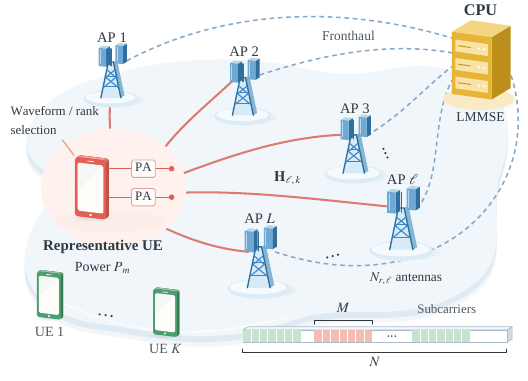}
\caption{Multiuser uplink with distributed reception. The highlighted UE with two radio-frequency (RF) chains selects the waveform and transmission rank. Solid links denote wireless channels, and dashed links denote fronthaul to the central processing unit for LMMSE equalization. The lower strip shows disjoint allocations of $M$ subcarriers within $N$ subcarriers. Wireless links for other UEs are omitted.}
\label{fig:system}
\end{figure}

Existing studies have investigated the power and equalization tradeoffs between CP-OFDM and DFT-s-OFDM \cite{wu2009,zhao2015,shao2022semantic}, waveform switching \cite{villena2025}, and spatial multiplexing under distributed reception \cite{bjornson2020,wang2022}. However, how these factors jointly determine waveform and rank selection remains insufficiently understood. In particular, improved spatial separation does not necessarily affect the two decisions in the same way: while it enhances the benefit of rank-2 transmission, its effect on waveform preference depends on the frequency distribution of the improvement. Understanding this distinction requires characterizing both the channel-dependent switching conditions and the spatial structure created by distributed AP observations.

In this paper, we investigate the joint selection of CP-OFDM and DFT-s-OFDM at rank 1 and noncoherent rank 2 under common UE power constraints and centralized linear minimum mean-square error (LMMSE) reception. Our main contributions are summarized as follows:
\begin{itemize}[leftmargin=0.5cm]
\item We derive waveform-switching thresholds at both ranks and a rank-selection margin that jointly characterize the four-mode decision. We further identify how the rank-2 waveform threshold depends on the wideband channel structure in the low- and high-SNR regimes.

\item We decompose aggregate spatial separation into three nonnegative contributions arising from local AP responses. We show that improving spatial separation, at fixed port gains and transmit powers and without reducing separation on any subcarrier, increases the rank-selection margin, whereas its effect on the rank-2 waveform threshold depends on which subcarriers improve. Numerical results demonstrate that AP deployment can change the preferred waveform even without changing the preferred rank.
\end{itemize}

\section{System Model}\label{sec:model}
Consider an uplink with $K$ UEs and $L$ cooperating APs. The APs forward their observations to a central processing unit (CPU) for joint processing. The system uses $N$-point OFDM modulation, and each UE occupies $M$ contiguous subcarriers with disjoint allocations satisfying $KM\le N$. The orthogonal allocations allow us to analyze one representative UE under a fixed schedule, omitting its UE index. Fig.~\ref{fig:system} illustrates the distributed reception network with a representative UE highlighted.

\subsection{Uplink Transmission}\label{sec:transmit}
Each UE has two transmit radio-frequency (RF) chains, each connected to a separate power amplifier (PA) and antenna port. Rank 1 uses one of the two antenna ports, while rank 2 maps two independent layers separately to the two ports, following the noncoherent precoding structure of NR uplink MIMO \cite{ts38211,ts38214}. Only the selected PA is active at rank 1, whereas both PAs operate with equal layer powers at rank 2. The corresponding precoding vectors at rank 1 are proportional to $[1,0]^T$ and $[0,1]^T$, while the rank-2 precoding matrix is proportional to $\mathbf I_2$. Rank-2 transmission provides spatial multiplexing, whose gain depends on the receiver's ability to separate the two port responses.

Within each rank, the waveform choice affects both PA operation and receiver processing. CP-OFDM maps constellation symbols directly onto the allocated subcarriers, whereas DFT-s-OFDM applies an $M$-point unitary DFT separately to each active layer \cite{ts38211}. The latter changes the transmit signal envelope and requires inverse DFT despreading after frequency-domain equalization. Both waveforms are considered at ranks 1 and 2 using the same allocated subcarriers, cyclic prefix (CP), and port mappings. We denote rank-1 CP-OFDM and DFT-s-OFDM by C1 and D1, and their rank-2 counterparts by C2 and D2.

The rank determines the number of active PAs, while the waveform affects the output backoff required to accommodate its signal envelope. Thus the four modes may have different available transmit powers under the same UE power constraints. Let $P_{\rm PC}$ denote the total transmit power determined by uplink power control before applying the UE and PA output power limits. The UE power limit and the saturation output power of one PA are denoted by $P_{\rm UE,max}$ and $P_{\rm PA,max}$, respectively. For mode $m\in\{\mathrm{C1},\mathrm{D1},\mathrm{C2},\mathrm{D2}\}$, the number of active PAs is $n_m$, with $n_{\mathrm{C1}}=n_{\mathrm{D1}}=1$ and $n_{\mathrm{C2}}=n_{\mathrm{D2}}=2$. For a given finite-alphabet modulation and PA operating criterion, mode $m$ is assigned an output backoff $B_m$ dB, giving a maximum average output power of $P_{\rm PA,max}10^{-B_m/10}$ per active PA. Combining the active PA capabilities with uplink power control and the UE total output power limit \cite{ts38213,ts381011}, the available total transmit power is modeled as
\begin{equation}\label{eq:power}
P_m=\min\!\left\{P_{\rm PC},\,P_{\rm UE,max},\,
 n_mP_{\rm PA,max}10^{-B_m/10}\right\}.
\end{equation}
All powers in \eqref{eq:power} are in linear units. The total power $P_m$ covers the $M$ allocated subcarriers and is divided equally between the two active ports at rank 2.

\subsection{Distributed Reception}\label{sec:reception}
AP $\ell$ has $N_{r,\ell}$ receive antennas, giving $N_{\rm R}=\sum_{\ell=1}^L N_{r,\ell}$ in total. We assume that the APs are synchronized and forward their frequency-domain observations to the CPU without loss. With perfect channel knowledge, the CPU performs centralized LMMSE equalization \cite{bjornson2020,shao2024theory}.

The channel is assumed constant over each data block, with the CP long enough to cover the channel delay spread. On allocated subcarrier $k\in\{0,\ldots,M-1\}$, receiver noise is independent across antennas and subcarriers, with variance $\sigma^2$. Let $\mathbf H_{\ell,k}\in\mathbb C^{N_{r,\ell}\times2}$ denote the channel from the two UE ports to AP $\ell$, normalized by $\sqrt{M\sigma^2}$ to account for the total power shared across $M$ subcarriers. Stacking these matrices vertically gives $\mathbf H_k$. At rank 2 with total power $P$, the normalized received signal is
\begin{equation*}
\mathbf y_k=\sqrt{P/2}\,\mathbf H_k\mathbf x_k+\mathbf n_k,
\quad \mathbf n_k\sim\mathcal{CN}(\mathbf0,\mathbf I_{N_{\rm R}}),
\end{equation*}
where $\E[\mathbf x_k\mathbf x_k^H]=\mathbf I_2$. Each active port therefore transmits $P/(2M)$ per allocated subcarrier.

To compare the two waveforms under the same linear receiver, we examine the LMMSE error variances \cite{wang2022}. These variances determine the Gaussian-input rates under independent scalar decoding and reveal the effect of DFT despreading \cite{wu2009,zhao2015}. The aggregate channel Gram matrix collects the two port gains and the inner product of their channel responses:
\begin{equation*}
\mathbf G_k\triangleq\sum_{\ell=1}^L\mathbf H_{\ell,k}^H\mathbf H_{\ell,k}
=\begin{bmatrix}a_k&c_k\\c_k^*&b_k\end{bmatrix}.
\end{equation*}
Here $a_k$ and $b_k$ are the aggregate port gains, and $c_k$ is the inner product of their channel responses. At rank 1, one port is active. Defining $g_{1,k}\triangleq a_k$ and $g_{2,k}\triangleq b_k$, the LMMSE error variance for port $i$ at total power $P$ is $(1+Pg_{i,k})^{-1}$.

At rank 2, with per-port power $p=P/2$, the LMMSE error covariance is $\mathbf E_k(p)\triangleq(\mathbf I_2+p\mathbf G_k)^{-1}$. Its diagonal entries depend on the port gains and their spatial correlation. To quantify the separation of the two port responses, define
\begin{equation}\label{eq:defdet}
d_k\triangleq a_kb_k-|c_k|^2,\qquad
\eta_k\triangleq\frac{d_k}{a_kb_k}\in[0,1],
\end{equation}
for $a_kb_k>0$. Aligned port responses give $\eta_k=0$, whereas orthogonal responses give $\eta_k=1$. The resulting error variances are
\begin{equation}\label{eq:eclosed}
e_{1,k}(p)=\frac{1+pb_k}{\det(\mathbf I_2+p\mathbf G_k)},\quad
e_{2,k}(p)=\frac{1+pa_k}{\det(\mathbf I_2+p\mathbf G_k)}.
\end{equation}
Since $\det(\mathbf I_2+p\mathbf G_k)=1+p(a_k+b_k)+p^2d_k$, greater spatial separation reduces both error variances at fixed port gains and $p>0$.

For CP-OFDM, the equalized symbols retain their individual subcarrier error variances. DFT-s-OFDM applies inverse DFT despreading after equalization. The unitary transform gives every recovered symbol on layer $i$ the average error variance $\bar e_i(p)\triangleq M^{-1}\sum_{k=0}^{M-1}e_{i,k}(p)$. Here and throughout, the overbar denotes averaging over the $M$ allocated subcarriers.

\subsection{Achievable Rates}\label{sec:rates}
Under independent scalar decoding with Gaussian inputs, residual interference is treated as noise, and an LMMSE error variance $e$ yields the scalar rate $-\log_2 e$ bit/s/Hz. At rank 1, the transmit port is selected for the entire data block to maximize the rate of the chosen waveform. Let $R_m(P)$ denote the rate of mode $m$ at total transmit power $P\ge0$. Using the error variances derived above, the four rates are
\begin{equation}\label{eq:fourrates}
\begin{aligned}
R_{\mathrm{C1}}(P)
 &=\max_{i=1,2}\overline{\log_2(1+Pg_{i,k})},\\
R_{\mathrm{D1}}(P)
 &=\max_{i=1,2}-\log_2\!\left[\overline{(1+Pg_{i,k})^{-1}}\right],\\
R_{\mathrm{C2}}(P)
 &=-\sum_{i=1}^2\overline{\log_2 e_{i,k}(P/2)},\\
R_{\mathrm{D2}}(P)
 &=-\sum_{i=1}^2\log_2\bar e_i(P/2).
\end{aligned}
\end{equation}
The rates are normalized per allocated subcarrier, with the common CP overhead omitted. Each mode is subject to $0\le P\le P_m$, where $P_m$ is given by \eqref{eq:power}. The four modes are compared at their respective available powers through $R_m(P_m)$.

\section{Mode Selection Analysis}\label{sec:gates}
Given the aggregate channel and the available transmit power of each mode, selecting among the four modes involves two related comparisons: waveform selection within each rank and a comparison between the best-performing modes at the two ranks. We establish the resulting switching conditions and identify the wideband channel properties that govern them.

\subsection{Waveform Switching Thresholds}
At a fixed rank, waveform selection can be expressed through the transmit power DFT-s-OFDM needs to match CP-OFDM at a given reference power. We quantify the rate difference at equal transmit power before expressing the waveform switching boundary as a power ratio.

From \eqref{eq:fourrates}, CP-OFDM averages the rates across subcarriers, whereas DFT-s-OFDM computes the rates after averaging the equalization errors. Jensen's inequality gives $R_{\mathrm Cr}(P)\ge R_{\mathrm Dr}(P)$ \cite{wu2009,zhao2015}. Define the rate gap at a common power as $\mathcal J_r(P)\triangleq R_{\mathrm Cr}(P)-R_{\mathrm Dr}(P)\ge0$. For rank 2, the gap takes the explicit form
\begin{equation}\label{eq:jensen}
\mathcal J_2(2p)=\sum_{i=1}^2
[\log_2\bar e_i(p)-\overline{\log_2 e_{i,k}(p)}].
\end{equation}
At rank 1, the same inequality holds for either port and is preserved when each waveform selects its best port over the entire block.

At the powers available under \eqref{eq:power}, the rate difference becomes
\begin{align*}
\Delta_r&\triangleq R_{\mathrm Dr}(P_{\mathrm Dr})-R_{\mathrm Cr}(P_{\mathrm Cr})\\
&=R_{\mathrm Cr}(P_{\mathrm Dr})-R_{\mathrm Cr}(P_{\mathrm Cr})
-\mathcal J_r(P_{\mathrm Dr}).
\end{align*}
The first difference measures the rate gained from the available transmit power, whereas the second is the rate lost at that same power. A nonnegative $\Delta_r$ favors DFT-s-OFDM.

To express this comparison as a power requirement, fix $P_{\mathrm Cr}>0$ and assume that the channel has at least one nonzero response. The rate $R_{\mathrm Dr}(P)$ is continuous and strictly increasing in $P$, but can be bounded in channels with spectral nulls or rank deficiency. Let $T_r\ge P_{\mathrm Cr}$ be the smallest finite power satisfying $R_{\mathrm Dr}(T_r)=R_{\mathrm Cr}(P_{\mathrm Cr})$, and set $T_r=+\infty$ if no such power exists. With $q_r\triangleq T_r/P_{\mathrm Cr}$, let $\gamma_r\triangleq P_{\mathrm Dr}/P_{\mathrm Cr}$ denote the available power ratio and let $\mathcal W_r^\star\triangleq\argmax_{w\in\{\mathrm C,\mathrm D\}}R_{wr}(P_{wr})$ denote the set of preferred waveforms at rank $r$. The selection rule is
\begin{equation}\label{eq:wavecondition}
\mathcal W_r^\star=
\begin{cases}
\{\mathrm C\}, & \gamma_r<q_r,\\
\{\mathrm C,\mathrm D\}, & \gamma_r=q_r,\\
\{\mathrm D\}, & \gamma_r>q_r,
\end{cases}
\qquad r\in\{1,2\}.
\end{equation}
The available power ratio $\gamma_r$ is governed by the UE and PA constraints, while $q_r\ge1$ is determined by the channel and the CP-OFDM reference power. If $q_r=+\infty$, CP-OFDM is strictly preferred at every finite $\gamma_r$. Equal available powers cannot give DFT-s-OFDM a strict rate advantage. At rank 1, $q_1=1$ precisely when at least one port maximizing $R_{\mathrm C1}(P_{\mathrm C1})$ has constant gain $g_{i,k}$ across the subcarriers. At rank 2, \eqref{eq:jensen} gives $q_2=1$ precisely when both error profiles $e_{i,k}(P_{\mathrm C2}/2)$ are constant.

For rank 2, the channel quantities governing the required power change with SNR. At low SNR, the leading terms depend on the variation of the aggregate port gains $a_k$ and $b_k$. At high SNR, suppressing the other stream leaves effective gains $a_k\eta_k$ and $b_k\eta_k$. These combine the port gains with their spatial separation. In the following, $\operatorname{Var}_k(x_k)\triangleq\overline{x_k^2}-(\bar x)^2$, and $\operatorname{GM}_k$ and $\operatorname{HM}_k$ denote the geometric and harmonic means over the allocated subcarriers.
\begin{theorem}[Rank-2 wideband switching requirement]\label{thm:wideband}
For a fixed wideband channel with positive definite $\mathbf G_k$ on every allocated subcarrier, let $P_{\mathrm C2}=2p$ and write $q_2(p)\triangleq T_2/(2p)$. Then
\begin{equation}\label{eq:widebandlimits}
\begin{aligned}
&q_2(p)=1+\frac{p}{2}\frac{\operatorname{Var}_k(a_k)+\operatorname{Var}_k(b_k)}{\bar a+\bar b}
+O(p^2),\quad p\to0,\\
&\lim_{p\to\infty}q_2(p)
=\left[\frac{\operatorname{GM}_k(a_k\eta_k)}{\operatorname{HM}_k(a_k\eta_k)}
\frac{\operatorname{GM}_k(b_k\eta_k)}{\operatorname{HM}_k(b_k\eta_k)}\right]^{1/2}.
\end{aligned}
\end{equation}
\end{theorem}
\begin{NewProof}
Using \eqref{eq:eclosed}, the two rank-2 errors have the low-SNR expansions $e_{1,k}(p)=1-pa_k+O(p^2)$ and $e_{2,k}(p)=1-pb_k+O(p^2)$. Taylor expansion of \eqref{eq:jensen} gives
\begin{equation*}
\mathcal J_2(2p)=\frac{p^2}{2\ln2}
[\operatorname{Var}_k(a_k)+\operatorname{Var}_k(b_k)]+O(p^3).
\end{equation*}
Writing $T_2=2p q_2(p)$, the leading rate increase from $q_2(p)-1$ is $p(\bar a+\bar b)(q_2(p)-1)/\ln2$. Equating it to the loss proves the first result. As $p\to\infty$, positive definiteness gives $e_{1,k}(p)\sim1/(p a_k\eta_k)$ and $e_{2,k}(p)\sim1/(p b_k\eta_k)$. Consequently,
\begin{align*}
R_{\mathrm C2}(2p)&=2\log_2p+\log_2\operatorname{GM}_k(a_k\eta_k)\\
&\quad+\log_2\operatorname{GM}_k(b_k\eta_k)+o(1),\\
R_{\mathrm D2}(2pq)&=2\log_2(pq)+\log_2\operatorname{HM}_k(a_k\eta_k)\\
&\quad+\log_2\operatorname{HM}_k(b_k\eta_k)+o(1).
\end{align*}
Equating the two rates yields the high-SNR limit.
\end{NewProof}

Theorem~\ref{thm:wideband} identifies two distinct wideband effects. Variation in port gains determines the first-order low-SNR power requirement. At high SNR, the requirement is governed by the frequency distribution of the effective gains after interference suppression and equals one precisely when both gain profiles are constant. To isolate the contribution of spatial separation, consider channels whose aggregate port gains do not vary with frequency.
\begin{corollary}[Loss from frequency-varying spatial separation]\label{cor:spatialloss}
Suppose $a_k=a>0$ and $b_k=b>0$ on every allocated subcarrier, while $0<d_k\le ab$ may vary. Then $q_1=1$ for every positive reference power. At rank 2, $q_2(p)>1$ for every $p>0$ if and only if $\operatorname{Var}_k(d_k)>0$. Moreover,
\begin{equation}\label{eq:flatgainlimits}
\begin{aligned}
&q_2(p)=1+\frac{p^3\operatorname{Var}_k(d_k)}{a+b}+O(p^4),\quad p\to0,\\
&\lim_{p\to\infty}q_2(p)=\frac{\operatorname{GM}_k(d_k)}{\operatorname{HM}_k(d_k)}.
\end{aligned}
\end{equation}
\end{corollary}
\begin{NewProof}
For rank 1, both port gain profiles are constant, so $q_1=1$. At rank 2, \eqref{eq:eclosed} shows that both errors are positive constants times $[1+p(a+b)+p^2d_k]^{-1}$. Thus their Jensen loss is positive exactly when $d_k$ varies. Expanding this loss gives $\mathcal J_2(2p)=p^4\operatorname{Var}_k(d_k)/\ln2+O(p^5)$, while the leading rate increase from $q_2-1$ is $p(a+b)(q_2-1)/\ln2$. This gives the low-SNR expansion. Substituting $a_k\eta_k=d_k/b$ and $b_k\eta_k=d_k/a$ into Theorem~\ref{thm:wideband} gives the high-SNR limit.
\end{NewProof}

Corollary~\ref{cor:spatialloss} separates the waveform decisions at the two ranks. With constant port gains, the C1--D1 threshold stays at one, while the C2--D2 boundary changes when spatial separation varies across frequency. Its low-SNR departure from one is of order $p^3$, and its high-SNR limit exceeds one whenever $d_k$ varies. Thus frequency-varying spatial separation can change the waveform preference at rank 2 without changing it at rank 1.

\subsection{Four-Mode Competition}
The waveform thresholds determine the preferred waveform at each rank, but the four-mode decision also requires comparing the two ranks. Let $C_r\triangleq R_{\mathrm Cr}(P_{\mathrm Cr})$ and $D_r\triangleq R_{\mathrm Dr}(P_{\mathrm Dr})$. The rank selection margin between the best waveform at rank 2 and the best waveform at rank 1 is
\begin{equation}\label{eq:rankmargin}
\begin{aligned}
\Lambda&\triangleq\max\{C_2,D_2\}-\max\{C_1,D_1\}\\
&=C_2-C_1+[\Delta_2]_+-[\Delta_1]_+,
\end{aligned}
\end{equation}
where $[x]_+\triangleq\max\{x,0\}$. A positive $\Lambda$ selects rank 2, while a negative $\Lambda$ selects rank 1. Let $m^\star$ be any mode maximizing $R_m(P_m)$ over the four modes. Combining this rank comparison with \eqref{eq:wavecondition}, the unique optimal mode, when there is no tie, is
\begin{equation}\label{eq:fourmode}
 m^\star=\begin{cases}
 \mathrm{C1},&\Lambda<0,\ \gamma_1<q_1,\\
 \mathrm{D1},&\Lambda<0,\ \gamma_1>q_1,\\
 \mathrm{C2},&\Lambda>0,\ \gamma_2<q_2,\\
 \mathrm{D2},&\Lambda>0,\ \gamma_2>q_2.
 \end{cases}
\end{equation}
At equality, all modes attaining the maximum rate are optimal. The rank boundary is $\Lambda=0$. On either side, only the waveform threshold $\gamma_r=q_r$ for the preferred rank determines the final choice.

The channel affects these comparisons through distinct quantities. At a fixed CP-OFDM reference power, $q_1$ and the rank-1 rates depend on the port gain profiles $a_k,b_k$, while rank-2 rates also depend on spatial separation $\eta_k$. With port gains and all four available powers fixed, spatial separation can affect $\Lambda$ and $q_2$ in \eqref{eq:fourmode}, while $q_1$ and the power ratios $\gamma_1,\gamma_2$ remain unchanged.
\begin{proposition}[Rank ordering under fixed port gains]\label{prop:rankboundary}
Consider positive semidefinite wideband Gram matrices with $a_kb_k>0$ and fixed positive mode powers. If spatial separation increases on one or more subcarriers and does not decrease on any other, the rank selection margin $\Lambda$ in \eqref{eq:rankmargin} strictly increases. Consequently, such an improvement can change the preferred rank from 1 to 2, but not from 2 to 1.
\end{proposition}
\begin{NewProof}
At fixed $a_k,b_k$, an increase in $\eta_k$ increases $d_k=a_kb_k\eta_k$ and thus $\det(\mathbf I_2+p\mathbf G_k)=1+p(a_k+b_k)+p^2d_k$ for every $p>0$. Both LMMSE errors in \eqref{eq:eclosed} strictly decrease on each improved subcarrier and do not increase elsewhere. Hence both rank-2 rates strictly increase at their fixed positive powers. Rank-1 rates are unchanged, so their maximization over the two waveforms preserves the strict increase in $\Lambda$.
\end{NewProof}

Proposition~\ref{prop:rankboundary} establishes how spatial separation orders the ranks, but the point at which rank 2 becomes preferable also depends on their available powers. A useful benchmark is a channel with constant port gains and spatial separation over frequency. There, CP-OFDM and DFT-s-OFDM have equal rates at the same power within each rank, and the four-mode decision reduces to a rank comparison with a closed-form separation threshold.
\begin{corollary}[Rank boundary with constant gains and separation]\label{cor:rankboundary}
Suppose $a_k=a>0$, $b_k=b>0$, and $\eta_k=\eta$ for every subcarrier. Let $P_1\triangleq\max\{P_{\mathrm C1},P_{\mathrm D1}\}$ and $P_2\triangleq\max\{P_{\mathrm C2},P_{\mathrm D2}\}$, and set $g_{\max}\triangleq\max\{a,b\}$ and $p\triangleq P_2/2$. Then rank 2 is strictly preferred if and only if $\eta>\eta_{\rm crit}$, where
\begin{equation}\label{eq:flatboundary}
\begin{split}
\eta_{\rm crit}\triangleq\frac{1}{p^2ab}\big[&
\sqrt{(1+P_1g_{\max})(1+pa)(1+pb)}\\
&-1-p(a+b)\big].
\end{split}
\end{equation}
\end{corollary}
\begin{NewProof}
Under the stated conditions, both waveforms have equal rates at a common power because the LMMSE error variances are constant across subcarriers. The best rank-1 rate is $\log_2(1+P_1g_{\max})$. At rank 2, \eqref{eq:eclosed} gives the best rate as $2\log_2[1+p(a+b)+p^2ab\eta]-\log_2[(1+pa)(1+pb)]$. Comparing these two rank rates and solving for $\eta$ gives \eqref{eq:flatboundary}.
\end{NewProof}
If $\eta_{\rm crit}<0$, rank 2 wins for every feasible $\eta\in[0,1]$. If $\eta_{\rm crit}\ge1$, rank 1 wins except for a possible tie at $\eta=1$. The threshold quantifies how much separation is needed to overcome differences in available transmit power and port gains.

These results distinguish two effects of spatial separation on the four-mode decision. At fixed port gains and available powers, increasing separation on at least one subcarrier without reducing it elsewhere strictly raises the rank selection margin $\Lambda$, while the C2--D2 threshold $q_2$ can shift in either direction. With constant port gains, Corollary~\ref{cor:spatialloss} further shows that frequency-varying separation introduces rank-2 DFT despreading loss without affecting the rank-1 waveform threshold.
\section{Effects of Distributed Reception}\label{sec:mechanisms}
Sec.~\ref{sec:gates} establishes the four-mode selection rule for a given aggregate channel. Under distributed reception, the placement of receive antennas across APs shapes the local channel responses and hence the aggregate port gains and spatial separability, including their variation across subcarriers. Connecting receiver deployment to mode selection therefore requires identifying how local AP responses form spatial separation. At fixed port gains and available powers, Proposition~\ref{prop:rankboundary} shows that increasing separation without reducing it on any subcarrier raises the rank selection margin, whereas the rank-1 rates in \eqref{eq:fourrates} leave the C1--D1 choice unchanged. This motivates a closer examination of how improvements at different subcarriers shift the C2--D2 waveform threshold.

\subsection{Sources of Spatial Separation}\label{sec:geometry}
Aggregate spatial separability reflects both local port correlations within APs and differences in their relative port responses across APs. We express these contributions through local Gram matrices.

Write $\mathbf G_{\ell,k}\triangleq\mathbf H_{\ell,k}^H\mathbf H_{\ell,k}$ with diagonal entries $a_{\ell,k}$, $b_{\ell,k}$ and off-diagonal entry $c_{\ell,k}$. For $a_kb_k>0$, define
\begin{equation*}
v_{\ell,k}\triangleq\sqrt{\frac{a_{\ell,k}b_{\ell,k}}{a_kb_k}},\qquad
c_{\ell,k}=\sqrt{a_{\ell,k}b_{\ell,k}}\,
\xi_{\ell,k}e^{\mathrm j\theta_{\ell,k}}.
\end{equation*}
Here $v_{\ell,k}$ weights the local port gains relative to their aggregate values, $0\le\xi_{\ell,k}\le1$ is the magnitude of the normalized local port correlation, and $\theta_{\ell,k}$ is its phase. Set $\theta_{\ell,k}=0$ whenever $c_{\ell,k}=0$. If either local port gain is zero, also take $v_{\ell,k}=\xi_{\ell,k}=0$.

\begin{proposition}[Sources of spatial separation]\label{prop:threeparts}
For any distribution of receive antennas among the APs, the normalized spatial separation admits the nonnegative decomposition
\begin{equation}\label{eq:threeparts}
\eta_k=\eta_{\mathrm{prof},k}+\eta_{\mathrm{loc},k}
+\eta_{\mathrm{cross},k},
\end{equation}
where
\begin{align*}
\eta_{\mathrm{prof},k}&\triangleq 1-\Big(\sum_\ell v_{\ell,k}\Big)^2,\\
\eta_{\mathrm{loc},k}&\triangleq\Big(\sum_\ell v_{\ell,k}\Big)^2
-\Big(\sum_\ell v_{\ell,k}\xi_{\ell,k}\Big)^2,\\
\eta_{\mathrm{cross},k}&\triangleq 4\sum_{\ell<\ell'}
 v_{\ell,k}v_{\ell',k}\xi_{\ell,k}\xi_{\ell',k}
 \sin^2\!\frac{\theta_{\ell,k}-\theta_{\ell',k}}{2}.
\end{align*}
\end{proposition}
\begin{NewProof}
Since $a_k=\sum_\ell a_{\ell,k}$ and $b_k=\sum_\ell b_{\ell,k}$, Cauchy-Schwarz gives $\sum_\ell v_{\ell,k}\le1$. Together with $0\le\xi_{\ell,k}\le1$, this establishes the nonnegativity of the first two terms, while the last term is manifestly nonnegative. Moreover,
\begin{equation*}
\eta_k=1-\frac{|c_k|^2}{a_kb_k}
=1-\Big|\sum_\ell v_{\ell,k}\xi_{\ell,k}
 e^{\mathrm j\theta_{\ell,k}}\Big|^2.
\end{equation*}
Expanding the squared magnitude and using $1-\cos x=2\sin^2(x/2)$ establishes \eqref{eq:threeparts}. An equivalent view follows from Cauchy-Binet: if $h_{ui,k}$ is row $u$, column $i$ of the stacked channel $\mathbf H_k$, then
\begin{equation*}
d_k=\sum_{1\le u<v\le N_{\rm R}}
|h_{u1,k}h_{v2,k}-h_{v1,k}h_{u2,k}|^2.
\end{equation*}
Each summand measures the ability of a pair of receiver observations to distinguish the two ports.
\end{NewProof}

The three terms reveal how the AP observations contribute to spatial separation:
\begin{itemize}[leftmargin=0.5cm]
\item \emph{Relative port gains across APs} ($\eta_{\mathrm{prof},k}$). The relative strengths of the two ports can differ from one AP to another. Such differences produce spatial separation even if the two port responses are aligned within every AP. This term vanishes when the two port gain profiles across APs are proportional.
\item \emph{Local separation} ($\eta_{\mathrm{loc},k}$). An AP with multiple receive antennas can distinguish the two port responses through their local spatial directions. This contribution is positive when an AP has nonzero gains for both ports and $\xi_{\ell,k}<1$, and vanishes when all contributing local responses are aligned.
\item \emph{Phases of local port correlations} ($\eta_{\mathrm{cross},k}$). Differences in these phases reduce the magnitude of the aggregate port correlation. They can therefore separate locally aligned responses even when the relative port gains are the same across APs.
\end{itemize}
For single-antenna APs, $\xi_{\ell,k}=1$ whenever both local port gains are nonzero. The local contribution then vanishes, while differences in relative port gains and differences in the phases of local port correlations can still provide spatial separation. These contributions share the same local channel parameters and may change together when the AP layout changes.

The decomposition holds separately on each subcarrier. Local port gains and correlations can vary with frequency, so the three contributions need not produce the same spatial separation at every $k$. Differences in propagation delays that multiply both port responses at an AP by a common phase do not create such variation: the phase cancels in $\mathbf H_{\ell,k}^H\mathbf H_{\ell,k}$. Frequency variation instead arises from changes in local port gains or relative port responses, as can occur through multipath. Thus AP observations determine both the amount of separation and the subcarriers on which it is available.

\subsection{Waveform Threshold Sensitivity}\label{sec:spatialgradient}
Contributions from individual APs to spatial separation need not be uniform across frequency. At fixed port gains, improving separation on a subcarrier lowers both rank-2 LMMSE errors there, but CP-OFDM averages subcarrier rates, whereas DFT-s-OFDM computes each layer's rate from its mean error after despreading. Consequently, the change in the C2--D2 switching threshold depends on which subcarrier improves.

We compare their responses at $P_{\mathrm C2}$ and $T_2$, where C2 and D2 achieve the same rate. If D2 gains more rate after a small improvement, less power is needed to match C2 and $q_2$ decreases. If C2 gains more, $q_2$ increases. Since $d_k=a_kb_k\eta_k$ at fixed port gains, $\partial q_2/\partial\eta_k=a_kb_k\,\partial q_2/\partial d_k$ has the same sign. We use the determinant to express this sensitivity.

\begin{proposition}[Waveform threshold sensitivity]\label{prop:qgradient}
Let $P_{\mathrm C2}>0$ and suppose $\mathbf G_k$ is positive definite on every allocated subcarrier. Vary one determinant $d_k$ in $0<d_k<a_kb_k$, holding all port gains, the other determinants, and $P_{\mathrm C2}$ fixed. Define $p_C\triangleq P_{\mathrm C2}/2$, $p_T\triangleq T_2/2$, and $f_k(p)\triangleq1+p(a_k+b_k)+p^2d_k$. Then
\begin{equation}\label{eq:qgradient}
\frac{\partial q_2}{\partial d_k}
=\frac{\mathcal A_k-\mathcal B_k}
{MP_{\mathrm C2}\ln2\,R_{\mathrm D2}'(T_2)},
\end{equation}
where
\begin{equation*}
\mathcal A_k\triangleq\frac{2p_C^2}{f_k(p_C)},\qquad
\mathcal B_k\triangleq\frac{p_T^2}{f_k(p_T)}
\sum_{i=1}^{2}\frac{e_{i,k}(p_T)}{\bar e_i(p_T)}.
\end{equation*}
The quantities $\mathcal A_k$ and $\mathcal B_k$ are proportional to the C2 and D2 rate responses at these powers. In particular,
\begin{equation*}
\sum_{i=1}^{2}\frac{e_{i,k}(p_T)}{\bar e_i(p_T)}\ge2
\quad\Longrightarrow\quad
\frac{\partial q_2}{\partial d_k}\le0.
\end{equation*}
\end{proposition}
\begin{NewProof}
By \eqref{eq:eclosed}, $e_{i,k}(p)$ is proportional to $f_k(p)^{-1}$ when the port gains are fixed. Differentiating the two rank-2 rates at fixed power gives
\begin{equation*}
\frac{\partial R_{\mathrm C2}(P_{\mathrm C2})}{\partial d_k}
=\frac{\mathcal A_k}{M\ln2},\quad
\left.\frac{\partial R_{\mathrm D2}(P)}{\partial d_k}\right|_{P=T_2}
=\frac{\mathcal B_k}{M\ln2}.
\end{equation*}
Implicit differentiation of $R_{\mathrm D2}(T_2)=R_{\mathrm C2}(P_{\mathrm C2})$ yields
\begin{equation*}
R_{\mathrm D2}'(T_2)\frac{\partial T_2}{\partial d_k}
=\frac{\mathcal A_k-\mathcal B_k}{M\ln2}.
\end{equation*}
Using $q_2=T_2/P_{\mathrm C2}$ proves \eqref{eq:qgradient}. For the sufficient condition, $q_2\ge1$ gives $p_T\ge p_C$. Since
\begin{equation*}
\frac{p^2}{f_k(p)}=\frac{1}{p^{-2}+(a_k+b_k)p^{-1}+d_k}
\end{equation*}
is increasing in $p>0$, the stated error condition implies $\mathcal B_k\ge\mathcal A_k$. Finally, $R_{\mathrm D2}'(T_2)>0$ establishes the sign.
\end{NewProof}

The sufficient condition identifies subcarriers with relatively large contributions to the despread error variances. At these powers, improving separation there yields a rate gain for D2 at least as large as that for C2, giving a nonpositive local sensitivity of $q_2$. For other subcarriers, $\mathcal A_k-\mathcal B_k$ determines the sign. The result therefore shows how the relative advantage of C2 and D2 responds to improvements at different frequency locations.

Proposition~\ref{prop:threeparts} connects the local AP responses to the spatial separation profile, while Proposition~\ref{prop:qgradient} identifies how improvements at individual subcarriers shift the C2--D2 threshold. Together with Proposition~\ref{prop:rankboundary}, they establish an asymmetry in the resulting mode changes. At fixed port gains and available powers, increasing separation without reducing it on any subcarrier can shift the preferred rank only toward rank 2. The C1--D1 choice remains unchanged, whereas the preferred waveform within rank 2 may switch in either direction. When an AP layout also changes the port gain profiles $a_k,b_k$, all four modes can be affected, and the preferred mode follows from \eqref{eq:fourmode}.
\section{Numerical Results}\label{sec:numerics}
Under a fixed receive-antenna budget, numerical results examine how local AP responses form spatial separation, how its variation across subcarriers affects switching boundaries, and how AP layouts jointly change the channel properties governing the four-mode decision.

\begin{figure}[!t]
\centering\includegraphics[width=0.8\columnwidth]{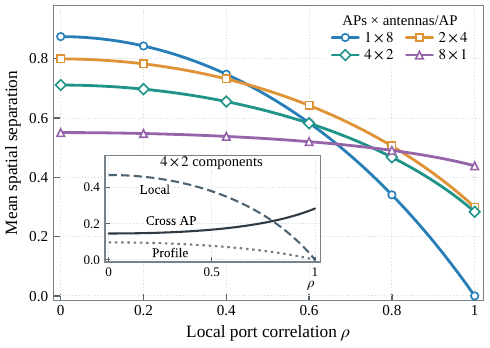}
\caption{Mean spatial separation across subcarriers and channel realizations versus the local port correlation magnitude $\rho$, with a fixed budget of eight receive antennas. The three components of \eqref{eq:threeparts} are also shown for the $4\times2$ layout.}
\label{fig:spatialnum}
\end{figure}

The UE has two transmit ports and eight receive antennas are allocated as $1\times8$, $2\times4$, $4\times2$, or $8\times1$ (APs $\times$ antennas per AP), with $1\times8$ as the single-AP reference. We use $N=1024$, $M=120$ contiguous subcarriers spaced by $30$ kHz, a $3.5$ GHz carrier, a noise spectral density of $-174$ dBm/Hz, and a $7$ dB receiver noise figure. With $P_{\rm PC}=26$ dBm, $P_{\rm UE,max}=23$ dBm, $P_{\rm PA,max}=24.5$ dBm, and output backoffs of $5$ dB for CP-OFDM and $2.5$ dB for DFT-s-OFDM at either rank, \eqref{eq:power} yields available powers of $19.5$, $22$, $22.51$, and $23$ dBm for C1, D1, C2, and D2, respectively. The DFT-s-OFDM power advantages are $2.5$ dB at rank 1 and $0.49$ dB at rank 2.

APs occupy a $300\times300$ m$^2$ area, with users uniformly distributed over $[20,280]^2$ m$^2$. The single AP is at $(150,150)$ m; $2\times4$ uses $(75,75)$ and $(225,225)$ m; $4\times2$ uses the four corners; and $8\times1$ adds the four edge midpoints. AP and UE heights are $10$ and $1.5$ m.

\subsection{Spatial Separation from AP Observations}
We first examine how AP observations form spatial separation using a three-tap channel with delays $(0,6,20)$ samples and relative powers $(0.65,0.25,0.10)$. At antenna $n$ of AP $\ell$, the two port coefficients of tap $t$ are proportional to $[z_{n,t},\,\rho e^{\mathrm j\phi_\ell}z_{n,t}+\sqrt{1-\rho^2}\,u_{n,t}]$, where $z_{n,t}$ and $u_{n,t}$ are independent $\mathcal{CN}(0,1)$ innovations across APs, antennas, and taps. We set $\phi_\ell=-\pi\cos\vartheta_\ell$, with $\vartheta_\ell$ the horizontal azimuth from the UE to AP $\ell$ relative to the $x$-axis. Channel attenuation follows the 3GPP urban microcell street canyon non-line-of-sight (UMi NLOS) path-loss model \cite{tr38901}, a minimum horizontal distance of $10$ m, and independent $4$ dB AP shadowing. The parameter $\rho\in[0,1]$ is the local port correlation magnitude. We vary it using $6400$ realizations with paired UE positions and Gaussian innovations across layouts and $\rho$.

\begin{figure}[!t]
\centering\includegraphics[width=0.85\columnwidth]{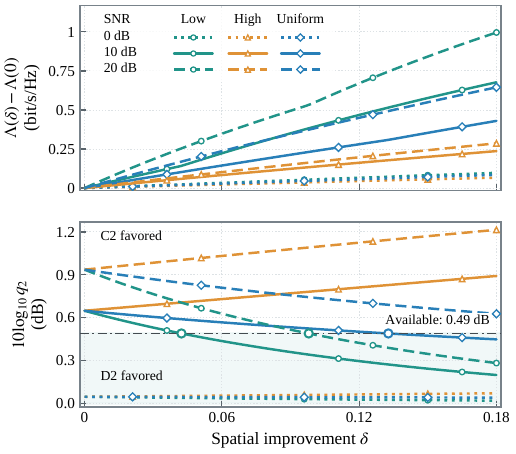}
\caption{Equal improvements in mean spatial separation yield different wideband mode responses. Top: change in the rank selection margin. Bottom: the required DFT-s-OFDM power ratio $q_2$, with the available rank-2 power advantage indicated by the horizontal line.}
\label{fig:widebandnum}
\end{figure}

At low correlation, the single AP benefits from eight observations with equal large-scale gain, while path loss and shadowing weight distributed observations unevenly. At $\rho=0.1$, the mean separation in Fig.~\ref{fig:spatialnum} is $0.866$ for $1\times8$ and $0.550$ for $8\times1$. The advantage reverses as the local port responses align: at $\rho=1$, separation vanishes for the single AP but remains $0.284$ for $4\times2$ and $0.438$ for $8\times1$. With equal local port gains and aligned responses, the profile and local terms in Proposition~\ref{prop:threeparts} vanish, leaving differences in the phases of local port correlations as the source of separation. The $4\times2$ decomposition illustrates this change in the dominant mechanism.

\subsection{Frequency Distribution of Spatial Separation}
To distinguish rank and waveform responses, we improve spatial separation at fixed port gains and mode powers. Following Corollary~\ref{cor:spatialloss}, we set $a_k=b_k=a$, with $\eta_k=0.2$ on half of the $M=120$ subcarriers and $\eta_k=0.8$ on the other half. For $\delta\in[0,0.18]$, we increase either half by $\delta$ or both halves by $\delta/2$, producing the same increase in mean separation. The reference SNR is $10\log_{10}(P_{\mathrm C2}a/2)$ dB in the normalized Gram model, evaluated at $0$, $10$, and $20$ dB with fixed ratios of the available mode powers.

\begin{figure}[!t]
\centering\includegraphics[width=0.8\columnwidth]{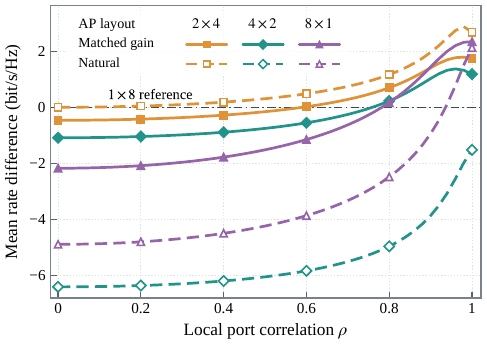}
\caption{Mean optimal rate difference between each distributed layout and the single-AP $1\times8$ reference. Natural propagation and comparisons at matched aggregate channel gain use the same four-mode selection rule.}
\label{fig:layoutnum}
\end{figure}

All three changes increase $\Lambda$ in Fig.~\ref{fig:widebandnum}, as predicted by Proposition~\ref{prop:rankboundary}. Since $\Lambda>0$ throughout, the optimal mode is within rank 2. At $10$ dB and $\delta=0.18$, improving the subcarriers with lower spatial separation reduces the required DFT power advantage from $0.648$ to $0.199$ dB, while improving the subcarriers with higher spatial separation raises it to $0.892$ dB. Against the available $0.49$ dB advantage, these changes favor D2 and C2, respectively. Subcarriers with weaker separation contribute more to the mean error variances of the DFT-s-OFDM layers, explaining the different threshold responses through Proposition~\ref{prop:qgradient}. Thus all three improvements favor rank 2, while their frequency locations can shift the C2--D2 boundary in opposite directions.

\subsection{Mode Selection Across AP Layouts}
Unlike the controlled changes above, AP layouts alter port gain profiles together with spatial separation, potentially changing the choice among all four modes. We compare \emph{natural propagation}, retaining each AP's path loss and shadowing, with \emph{matched aggregate channel gain}, scaling each channel realization to match the subcarrier-averaged Gram trace of the paired $4\times2$ layout. The latter preserves differences in relative port gains, spatial separation, and frequency profiles. Each realization selects the mode with the highest rate at its available power under the same eight-antenna budget.

Fig.~\ref{fig:layoutnum} shows that the layout ordering reverses as local port correlation rises, even with matched aggregate channel gain. The $4\times2$ layout trails the single AP by $0.98$ bit/s/Hz at $\rho=0.3$, but exceeds it by $1.19$ bit/s/Hz at $\rho=1$; $8\times1$ gains $2.35$ bit/s/Hz at $\rho=1$. The single AP benefits from strong local separation at low correlation, while observations from different APs can sustain two spatial streams when local responses align. Under natural propagation, differences in aggregate channel gain also affect the ordering, and $4\times2$ remains below the single-AP reference over the plotted range.

\begin{figure}[!t]
\centering\includegraphics[width=0.85\columnwidth]{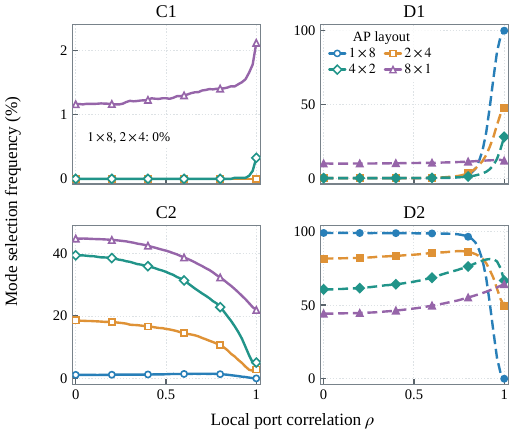}
\caption{Optimal selection frequencies of all four modes at matched aggregate channel gain. Top: C1 and D1. Bottom: C2 and D2. Colors denote the AP layouts, and line styles distinguish the waveforms.}
\label{fig:modenum}
\end{figure}

The selection frequencies in Fig.~\ref{fig:modenum} reveal changes in both rank and waveform. At $\rho=0.3$, rank 2 is preferred in nearly all realizations for both $1\times8$ and $4\times2$, yet the C2 share increases from $1.2\%$ to $37.3\%$. The preferred waveform can therefore change without a rank change. At $\rho=1$, the single AP always selects D1, whereas $8\times1$ selects all four modes, with D2 most frequent ($64.0\%$). Across the same realizations, adaptation also matters: for $2\times4$ at $\rho=1$ with matched aggregate channel gain, selecting among the four modes for each channel realization gains $1.75$ bit/s/Hz over the best fixed mode.
\section{Conclusion}\label{sec:conclusion}
This paper investigated the joint selection of uplink waveform and rank under distributed reception. The results reveal that the spatial organization of receive resources affects not only transmission performance but also the relative merits of different transmission modes. Even with the same receive antenna budget, different AP deployments can yield aggregate channels with spatial and frequency structures that favor different transmission strategies. This insight offers a broader design perspective for 6G distributed radio access by incorporating the spatial organization of receive resources into uplink mode adaptation and providing a theoretical basis for the joint design of transmission strategies and receive architectures.
\bibliographystyle{IEEEtran}
\bibliography{refs}
\end{document}